\documentclass[useAMS,usenatbib]{mn2e}

\usepackage{epsfig,graphics}
\usepackage{epsf  ,graphicx}

\newcommand{\Mwd  } {M_{\rm wd}}
\newcommand{\Rwd  } {R_{\rm wd}}

\newcommand{\Om   }   {\Omega}
\newcommand{\Odot }   {\dot{\Omega}}
\newcommand{\Oddot}   {\ddot{\Omega}}
\newcommand{\Pdot }   {\dot{P}}
\newcommand{\Pddot}   {\ddot{P}}

\newcommand{\ASCA}   {{\it ASCA\/}}
\newcommand{\XMM}    {{\it XMM}}
\newcommand{\XMMN}   {{\it XMM\/-Newton}}
\newcommand{\Chandra}{{\it Chandra\/}}

\title[The white dwarf in AE Aqr brakes harder]%
      {The white dwarf in AE Aqr brakes harder}

\author[C.~W.\ Mauche]
{Christopher W.\ Mauche\thanks{E-mail:
mauche@cygnus.llnl.gov}\\%
Lawrence Livermore National Laboratory,
L-473, 7000 East Avenue, Livermore, CA 94550}

\begin{document}

\date{Accepted 2006 April 13. Received 2006 April 12; in original form
2005 December 12}

\pagerange{\pageref{firstpage}--\pageref{lastpage}} \pubyear{2006}

\maketitle

\label{firstpage}

\begin{abstract}
Taking advantage of the very precise de Jager et al.\ optical white dwarf
orbit and spin ephemerides; \ASCA, \XMMN, and \Chandra\ X-ray observations
spread over 10 yrs; and a cumulative 27 yr baseline, we have found that in
recent years the white dwarf in AE Aqr is spinning down at a rate that is
slightly faster than predicted by the de Jager et al.\ spin ephemeris. At
the present time, the observed period evolution is consistent with either
a cubic term in the spin ephemeris with $\Pddot=3.46(56)\times 10^{-19}~\rm
d^{-1}$, which is inconsistent in sign and magnitude with magnetic-dipole
radiation losses, or an additional quadratic term with $\Pdot =2.0(1.0)
\times 10^{-15}~{\rm d~d^{-1}}$, which is consistent with a modest increase
in the accretion torques spinning down the white dwarf. Regular monitoring,
in the optical, ultraviolet, and/or X-rays, is required to track the
evolution of the spin period of the white dwarf in AE Aqr.
\end{abstract}



\section{Introduction}

AE Aqr is a bright ($V\approx 11$), nova-like cataclysmic binary consisting of
a magnetic white dwarf primary and a K4--5 V secondary with a long 9.88 hr
orbital period and the shortest known white dwarf spin period $P=33.08$ s
\citep{pat79}. Although originally classified and interpreted as a
disk-accreting DQ Her star \citep{pat94}, AE Aqr displays a number of unusual
features that are not naturally explained by this model. First, violent
flaring activity is observed in the radio, optical, ultraviolet, X-ray,
and TeV $\gamma$-rays. Second, the Balmer emission lines are single peaked and
produce Doppler tomograms that are not consistent with those of an accretion
disk. Third, the white dwarf is spinning down at a rate $\Pdot = 5.64 \times
10^{-14}~{\rm s~s^{-1}}$ \citep[hereafter, de Jager]{deJ94}. Although this
corresponds to the small rate of change of $1.78~\rm ns~yr^{-1}$, AE Aqr's
spin-down is typically characterized as ``rapid'' because the characteristic
time $P/\Pdot\approx 2\times 10^7$ yr, which is short compared to the lifetime
of the binary, and because the spin-down luminosity $L_{\rm sd}=-I\Om\Odot
\approx 1\times 10^{34}~\rm erg~s^{-1}$ (where $I\approx 0.2\Mwd\Rwd^2 \approx
2\times 10^{50}~\rm g~cm^2$ is the white dwarf moment of inertia, $\Mwd $ and
$\Rwd $ are the white dwarf mass and radius, respectively, $\Om = 2\pi/P$,
and $\Odot = -2\pi\Pdot/P^2$), which exceeds the secondary's thermonuclear
luminosity by an order of magnitude and the accretion luminosity by two orders
of magnitude. Given this, AE Aqr could be thought of as being powered primarily
by the ultimate in clean energy sources: a flywheel.

Because of its unique properties and variable emission across the
electromagnetic spectrum, AE Aqr has been the subject of numerous studies,
including an intensive multiwavelength observing campaign in 1993
October \citep[and the series of papers in ASPC 85]{cas96}. Based on these
studies, AE Aqr is now widely believed to be a former supersoft X-ray binary
\citep{sch02} and current {\it magnetic propeller\/} \citep{wyn97}, with most
of the mass lost by the secondary being flung out of the binary by the magnetic
field of the rapidly rotating white dwarf. These models explain many of AE
Aqr's unique characteristics, including the fast spin rate and rapid secular
spin-down rate of the white dwarf, the anomalous spectral type of the
secondary, the anomalous C to N abundance \citep{mau97}, the absence of
signatures of an accretion disk \citep{wel98}, the violent flaring activity
\citep{per03}, and the origin of the radio and TeV $\gamma$-ray emission
\citep{kui97, mei03}.

To build on this observational and theoretical work, while taking advantage
of a number of improvements in observing capabilities, during 2005 August
28--September 2 a group of professional and amateur astronomers conducted a
campaign of multiwavelength (radio, optical, ultraviolet, X-ray, and TeV
$\gamma$-ray) observations of AE Aqr. Analyses of these data are ongoing, but
here we present a fundamental result -- the spin period of the white dwarf
-- that relies solely on photometric data from the {\it Chandra X-ray
Observatory\/} and archival data from \ASCA\ and \XMMN .

\section{Observations and Analysis}

AE Aqr was observed by \ASCA\ with the Gas Imaging Spectrometer (GIS) detectors
and the Solid-State Imaging Spectrometer (SIS) detectors beginning on 1995
October 14 at 00:16 UT for 82 ks (Sequence Number 33005000),
by \XMM\ with the European Photon Imaging Collaboration (EPIC) pn CCD detector
beginning on 2001 November 7 at 23:47 UT for 14 ks (ObsID 0111180201),
and by
\Chandra\ with the High-Energy Transmission Grating (HETG) and the Advanced
CCD Imaging Spectrometer (ACIS) detector beginning on 2005 August 30 at 06:37
UT for 81 ks (ObsID 5431).
The \ASCA\ and \XMM\ data have been previously discussed by \cite{era99,cho99};
Osborne [2002, unpublished presentation at the Third Magnetic Cataclysmic
Variable Workshop (IAU Coll.\ 190)]; and \cite{ito06}.

The \ASCA\ and \XMM\ data and the \Chandra\ data were extracted from the High
Energy Astrophysics Science Archive Research Center and the \Chandra\ Data
Archive, respectively. The data files used for subsequent analysis were the
\ASCA\ GIS and SIS screened bright mode event files, the \XMM\ EPIC pn pipeline
processed event file, and the \Chandra\ level 2 pipeline processed event file.
The data in these files were manipulated as follows. First, all times were
converted from spacecraft time to Terrestrial Time (TT) and corrected to the
solar system barycenter using the \ASCA\ {\sc FTOOLS}\footnote{Available at
http://heasarc.gsfc.nasa.gov/docs/software/lhea\-soft/ftools/.} v6.0 tool {\tt
timeconv}, the \XMM\ Science Analysis Software ({\sc SAS}\footnote{Available at
http://xmm.vilspa.esa.es/sas/.}) v6.5.0 tool {\tt barycen}, and the \Chandra\
Interactive Analysis of Observations ({\sc CIAO}\footnote{Available at
http://cxc.harvard.edu/ciao/.}) v3.2 tool {\tt axbary}. The barycentric
corrections produced by these different software packages were checked against
those of the Interactive Data Language ({\sc IDL}) procedure
{\tt barycen}\footnote{Available at
http://astro.uni-tuebingen.de/software/idl/aitlib\-/astro/.} (which does not by
default account for the varying light travel time between the satellite and the
center of Earth), and were found to agree within 40, 210, and 200 ms for \ASCA,
\XMM, and \Chandra, respectively, consistent with the size of each satellite's
orbit. Second, source and background events were extracted from the event files
using custom IDL software. For \Chandra, events were collected from the
zeroth order image and the $\pm $ first order dispersed spectrum using the
region masks in the level 2 pha file. Third, event times $t$ were converted to
white dwarf orbit and spin phases via the relations
$\phi_{\rm orb} =\Omega_{\rm orb}(t-T_0)$ and
$\phi_{\rm spin}=\Om_0(t-T_{\rm max}) + {1\over2}\Odot(t-T_{\rm max})^2$,
respectively, where
$\Omega_{\rm orb}=2\pi/P_{\rm orb}$,
$\Om_0=2\pi/P_{33}$, and
$\Odot = -2\pi\Pdot_{33}/P_{33}^2$ and
$T_0$, $P_{\rm orb}$,
$T_{\rm max}({\rm BJD})\approx 2445172$,
$P_{33}=0.00038283263840$ d, and
$\Pdot_{33}=5.642\times 10^{-14}~\rm d~d^{-1}$
are the white dwarf orbit and spin ephemeris constants from Table 4 of de
Jager.\footnote{Note that the expression in \S 6 of de Jager for the times of
spin maxima formally introduces a term proportional to $(t-T_{\rm max})^3$
in the expression for $\phi_{\rm spin}$, but it amounts to an insignificant
$2\times 10^{-5}$ cycles at the time of the \Chandra\ observation.} Fourth,
filters were applied to restrict attention to events from two orbital cycles
for \ASCA\ and \Chandra\ and ${3\over 8}$ of an orbital cycle centered on
$\phi_{\rm orb}=0.25$ for \XMM. The resulting range of orbit phases and
observation dates are listed in Table 1. Fifth, background-subtracted spin
phase folded count rate light curves were calculated and fit with the cosine
function $A + B \cos (\phi_{\rm spin}- \phi_0)$, where $A$ is the mean count
rate, $B$ is the pulse semiamplitude, and $\phi_0$ is the phase offset
(expressed in cycles $\phi_0/2\pi$ in what follows). Given that the optical,
ultraviolet, and X-ray spin pulses are aligned in phase \citep{pat80,era95},
$\phi_0$ should be equal to zero to within the uncertainty of the de Jager spin
ephemeris, if the ephemeris remains valid at the times of the \ASCA, \XMM, and
\Chandra\ observations.

\begin{figure}
\label{fig1}
\includegraphics{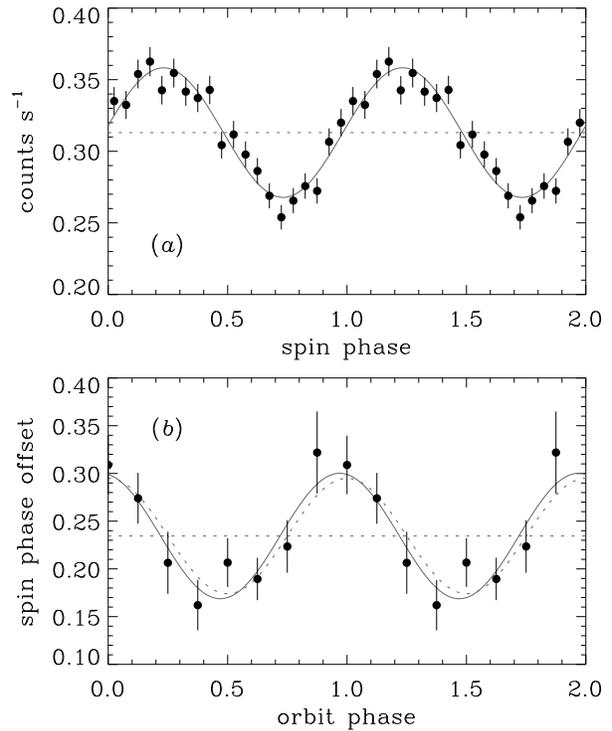}
\caption{({\it a\/}) \Chandra\ spin phase folded count rate light curve of AE
Aqr ({\it filled circles with error bars\/}), best-fitting cosine ({\it solid
curve)\/}, and mean count rate $A$ ({\it dotted line\/}). ({\it b\/}) Spin
phase offset as a function of orbit phase ({\it filled circles with error
bars\/}), best-fitting cosine ({\it solid curve)\/}, mean phase offset ({\it
dotted line\/}), and the predicted spin phase offset variation for the de
Jager white dwarf orbit ephemeris and a pulse time delay of 2 s ({\it dotted
curve\/}).}
\end{figure}

\begin{table*}
\centering
\begin{minipage}{140mm}
\caption{Log of observations.}
\begin{tabular}{lccc}
\hline
           & Orbit Phase               & Date                     \\
Satellite & $\phi_{\rm orb}$ (cycles) & [$\rm BJD(TT)-2400000$]  \\
\hline
\ASCA     & 11738.8000--11740.8000    & 50004.62127--50005.44459 \\
\XMM      & 17124.0625--17124.4375    & 52221.49479--52221.64916 \\
\Chandra  & 20504.0000--20506.0000    & 53612.86503--53613.68834 \\
\hline
\end{tabular}
\end{minipage}
\end{table*}

\subsection{\textbfit{Chandra} results, ignoring the pulse time delays}

We begin the analysis by considering the \Chandra\ data, which covers two full
binary orbits of AE Aqr without interruption by Earth occultations or
detector shutdowns. As shown in the top panel of Figure 1, the \Chandra\
background-subtracted spin phase folded count rate light curve of AE Aqr is
reasonably well fit ($\chi_\nu^2 =19.1/17 = 1.12$) by the cosine function with
a mean count rate $A=0.313\pm 0.002~\rm counts~s^{-1}$, a relative amplitude
$B/A=15\pm 1$ per cent, and a phase offset $\phi_0=0.232\pm 0.011$ cycles
(throughout the paper, errors are $1\sigma $ or 68 per cent confidence for 1
degree of freedom); the phase offset differs from the de Jager spin ephemeris
by $4.4\sigma $. To check if this result is affected by intensity variations
around the orbit, we repeated the above procedure for each of 8 contiguous
orbit phase intervals centered on $\phi_{\rm orb}=[0,1,2,\ldots, 7]/8$. The
resulting variation of the spin phase offset with orbit phase is shown in the
bottom panel of Figure 1 and is well fit ($\chi_\nu^2 = 4.31/5 = 0.86$) by the
cosine function $A' + B' \cos (\phi_{\rm orb}-\phi_0')$, with $A'=0.234\pm
0.010$ spin cycles, $B'=0.066\pm 0.015$ spin cycles, and $\phi_0'=-0.03\pm
0.03$ orbit cycles: the orbit phase offset $\phi_0'$ is consistent with zero,
the mean spin phase offset $A'$ is consistent with the value derived above for
$\phi_0$, and the semiamplitude $B'$ corresponds to a pulse time delay
$B'P_{33} =2.17\pm 0.48$ s, which is consistent with that measured in the
optical ($2.04\pm 0.13$ s, de Jager) and ultraviolet \cite[$1.93\pm 0.03$
s,][]{era94}. From this result,\footnote{\cite{rei95} previously reported that
the {\it ROSAT\/} X-ray pulse time delays vary with the orbital period and are
in phase with the optical/ultraviolet pulse time delays, but they did not
provide details.} we conclude that the source of the pulsating X-rays follows
the motion of the white dwarf as it orbits around the center of mass in AE Aqr.

\subsection{\textbfit{ASCA}, \textbfit{XMM}, and \textbfit{Chandra} results,
accounting for the pulse time delays}

Given these results, we analyzed the \ASCA, \XMM, and \Chandra\ data for AE
Aqr assuming that in each case the X-ray source follows the motion of the
white dwarf. Specifically, we corrected the barycentric event times by
$2\cos(\phi_{\rm orb})$ s before calculating the spin phases, the spin phase
folded light curves, and the cosine fit parameters. The pulse time delay
correction allows us to use all the data for each observation, without regard
to gaps due to Earth occultations or detector shutdowns (\ASCA ) or to the
varying source intensity (all three observations). With the pulse time delay
corrections and the previous orbit phase filters, the spin phase folded count
rate light curves and the best-fitting cosines are as shown in Figure 2 and the
best-fitting cosine parameters and the Barycentric Julian Dates of the X-ray
pulse maxima are as listed in Table 2. Figure 3 plots the phase offsets versus
time, demonstrating that the observed phases $O$ are diverging from the
calculated phases $C$ assuming the de Jager spin ephemeris: accounting for
the error in the ephemeris, the two differ by $1.7\sigma $, $3.9\sigma $, and
$4.4\sigma $ for \ASCA, \XMM, and \Chandra, respectively. Apparently, the white
dwarf in AE Aqr is slowly but progressively ``coming off the rails'' of the de
Jager spin ephemeris.

\begin{figure}
\label{fig2}
\includegraphics{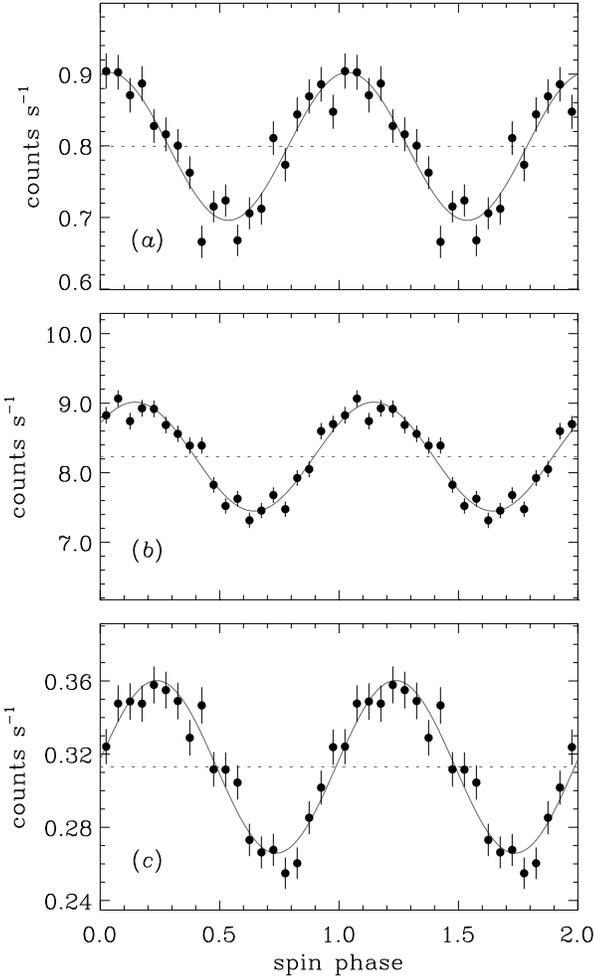}
\caption{({\it a\/}) \ASCA , ({\it b\/}) \XMM , and ({\it c\/}) \Chandra\ spin
phase folded count rate light curves of AE Aqr ({\it filled circles with error
bars\/}), best-fitting cosines ({\it solid curves)\/}, and mean count rates $A$
({\it dotted lines\/}). Each panel is scaled to $\pm 25$ per cent of the mean
count rate.}
\end{figure}

\begin{table*}
\centering
\begin{minipage}{140mm}
\caption{Best-fitting cosine parameters $A+B\cos(\phi_{\rm spin}-\phi_0)$.}
\begin{tabular}{lcccccc}
\hline
           &              & $A$                   & $B$ 
&       & $\phi_0$      &  Date of pulse maximum    \\
Satellite & $\chi_\nu^2$ & ($\rm counts~s^{-1}$) & ($\rm 
counts~s^{-1}$) & $B/A$ & (rel. cycles) & [$\rm BJD(TT)-2400000$]   \\
\hline
\ASCA     & 1.43 & $0.799\pm 0.005$ & $0.103\pm 0.007$ & $0.129\pm 
0.009$ & $ 0.037\pm 0.011$ & $50004.7037529\pm 0.0000044$ \\
\XMM      & 1.66 & $8.232\pm 0.026$ & $0.783\pm 0.037$ & $0.095\pm 
0.004$ & $ 0.146\pm 0.007$ & $52221.5720142\pm 0.0000029$ \\
\Chandra  & 1.09 & $0.313\pm 0.002$ & $0.047\pm 0.003$ & $0.151\pm 
0.009$ & $ 0.237\pm 0.010$ & $53613.2767058\pm 0.0000039$ \\
\hline
\end{tabular}
\end{minipage}
\end{table*}

\section{Updating the AE Aqr white dwarf spin ephemeris}

The trend displayed in Figure 3 of the difference between the observed and
calculated AE Aqr white dwarf spin phases is steeper than can be explained by
an increase in either the value of the $\Pdot_{33}$ parameter or the size of
the parameter uncertainties in the de Jager spin ephemeris. Given the limited
baseline and the limited amount of X-ray data, the observed trend is consistent
with the addition to the ephemeris of either a $\Pddot_{33}$ term centered on
$T_{\rm max}$ or an additional $\Pdot_{33}$ term centered on a late date
$T_{\rm max}'$. In the first case, the Taylor expansion of the spin frequency
$\Om(t)=\Om_0 + \Odot (t-T_{\rm max}) + {1\over 2}\Oddot (t- T_{\rm max})^2$
results in a cubic term in the expansion for
$\phi_{\rm spin} = \Om_0(t-T_{\rm max}) + {1\over2 } \Odot(t-T_{\rm max})^2
+ {1\over6}\Oddot(t-T_{\rm max})^3$ \citep{man77}, so
$O-C = -{1\over6} \Oddot(t-T_{\rm max})^3$
and the optical and X-ray data are consistent with
$\Oddot=-1.48(24)\times 10^{-11}~\rm d^{-3}$ or
$\Pddot_{33} = 3.46(56)\times 10^{-19}~\rm d^{-1}$. In the second case,
$O-C = -{1\over 2} \Odot' (t- T_{\rm max}')^2$
and the X-ray data are consistent with
$T_{\rm max}'({\rm BJD}) = 2447650\pm 1200$ and
$\Odot'=-8.5(4.4)\times 10^{-8}~\rm d^{-2}$ or
$\Pdot_{33}'=2.0(1.0)\times 10^{-15}~{\rm d~d^{-1}}$.
(In both cases, the quoted errors account for the error in the de Jager spin
ephemeris.) Of the two options, the quadratic fit is preferred, since it
provides a slightly better fit to the data and produces phase residuals that
are smaller ($\leq 0.07$ cycles for the quadratic term compared to $\leq 0.13$
for the cubic term) during the epoch studied by de Jager (1978.5--1992.6). In
either case, the frequency derivatives are negative and the period derivatives
are positive, so during the epoch covered by the \ASCA, \XMM, and \Chandra\
X-ray observations (1995.8--2005.7), the white dwarf in AE Aqr was braking
(slightly) harder than described by the de Jager spin ephemeris.

\begin{figure}
\label{fig3}
\includegraphics{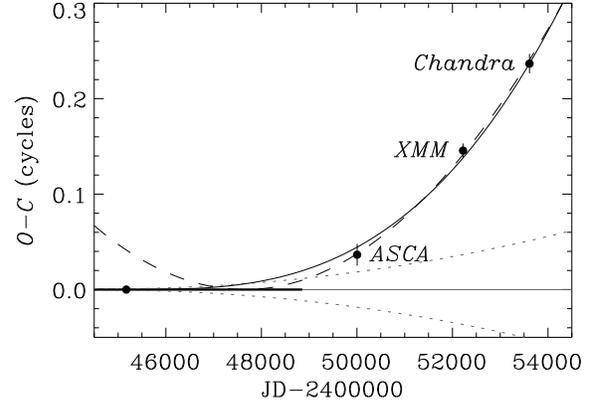}
\caption{Observed minus calculated spin phases of AE Aqr as a function of time
({\it filled circles with error bars\/}), the $1\sigma $ error envelope of the
de Jager spin ephemeris ({\it dotted curves\/}), the cubic fit to the de Jager
optical datum and the \ASCA, \XMM, and \Chandra\ X-ray data ({\it solid
curve\/}), and the quadratic fit to the X-ray data ({\it dashed curve\/}).
The epoch studied by de Jager is indicated by the thick horizontal line.}
\end{figure}

\section{Discussion}

Taking advantage of the very precise de Jager optical white dwarf orbit and
spin ephemerides; \ASCA, \XMM, and \Chandra\ X-ray observations spread over 10
yrs; and a cumulative 27 yr baseline, we have found that in recent years the
white dwarf in AE Aqr is spinning down at a rate that is slightly faster than
predicted by the de Jager spin ephemeris. At the present time, the observed
period evolution is consistent with either a cubic term in the spin ephemeris
with $\Pddot_{33}  =3.46(56)\times 10^{-19}~\rm d^{-1}$ centered on
$T_{\rm max}$ or an additional quadratic term with $\Pdot_{33}'=2.0(1.0)
\times 10^{-15}~{\rm d~d^{-1}}$ centered on $T_{\rm max}'({\rm BJD}) =
2447650\pm 1200$. We consider the implications of each option in turn.

If the observed $O-C$ residuals are due to a cubic term in the spin ephemeris,
it is possible to make an interesting comparison between the AE Aqr white
dwarf X-ray pulsar and neutron star radio pulsars. In a vacuum, a rotating
star with a misaligned magnetic dipole loses rotational kinetic energy via
magnetic-dipole radiation at a rate $L_{\rm sd}=-I\Om\Odot =
2\mu^2\sin^2\theta\Om^4/3c^3$, where $\mu=BR^3$ is the magnetic moment, $B$
is the surface magnetic field strength, $R$ is the stellar radius, and
$\theta$ is the angle between the rotation and magnetic axes. It is convenient
to express this as $\Odot = -K\Om^n$ with $K=2\mu^2\sin^2\theta/ 3Ic^3$ and
the so-called braking index $n=3$. Ignoring the fact that the white dwarf
in AE Aqr is not isolated [the light cylinder radius ($r_{\rm lc}=c/\Om =
1.6\times 10^{11}$ cm) is comparable to the binary separation ($a\approx
2\times 10^{11}$ cm), so the white dwarf magnetic field will drag against
the secondary, the magnetic field of the secondary, and the mass lost by the
secondary], if one assumes that the observed spin-down is due solely to
magnetic-dipole radiation losses, one obtains that $\mu\approx 1\times
10^{34}~\rm G~cm^3$ or $B\sim 50$ MG \citep{ikh98}, comparable to that
observed in polars. However, one also obtains $\Oddot= n\Odot^2/\Om =
1.07\times 10^{-15}~\rm d^{-3}$ or $\Pddot=-8.40\times 10^{-24}~\rm d^{-1}$,
which results is a {\it decrease\/} in $O-C$ by 0.1 cycles in $10^5$ yrs,
while we have observed an {\it increase\/} in $O-C$ by 0.24 cycles in 27 yrs.
Expressed another way, we have derived not $n=3$ but $n\approx -42000$.
Clearly, the enhanced spin-down of the white dwarf in AE Aqr has nothing to
do with magnetic-dipole radiation losses. On the other hand, if the observed
$O-C$ residuals are due to an additional quadratic term in the spin ephemeris,
the spin-down rate is a modest $3.5\pm 1.8$ per cent greater than the rate
derived by de Jager, which easily could be accommodated by, e.g., a small
increase in the mass transfer rate from the secondary, leading to an enhanced
spin-down torque on the white dwarf.

While we have found that additional low-order terms in the Taylor series
expansion of the white dwarf spin ephemeris adequately fit the trend of the
$O-C$ residuals of the \ASCA, \XMM, and \Chandra\ X-ray light curves of AE
Aqr, providing independent evidence of the nominal validity of the de Jager
ephemerides, we caution that our results are based on only three data points
spread over 10 yrs. First, it is possible that the spin evolution of AE Aqr
is more complex than assumed, that $O-C$ varies on shorter timescales and
manifests larger excursions than sampled, and that the good quadratic and
cubic fits to the $O-C$ residuals of the X-ray data are fortuitous. Second,
the quadratic and cubic spin ephemerides of \S 3 should be used with caution,
since they have yet to be shown to have any {\it predictive\/} capability.
Regular monitoring, in the optical, ultraviolet, and/or X-rays, is required
to track the evolution of the AE Aqr pulse period, to determine if the white
dwarf continues to spin down at the current rate, or if (as seems likely) it
varies stochastically in response to changes in the mass-transfer rate, the
varying efficiency of the magnetic propeller (e.g., the fraction of the
mass-transfer rate that is expelled from the binary versus that accreted by
the white dwarf), etc.

\section*{Acknowledgments}

The author is pleased to acknowledge Okkie de Jager for an educational and
entertaining exchange of e-mails, Koji Mukai for resolving a problem with
the \ASCA\ {\sc FTOOL} {\tt barycen}, Mike Arida for help installing the
\XMM\ {\sc SAS}, and Klaus Reinsch, Bill Welsh, Nazar Ikhsanov, and the
anonymous referee for information, comments, and suggestions that improved
the quality and clarity of this manuscript. This research made use of \ASCA\
and \XMM \ data obtained from the High Energy Astrophysics Science Archive
Research Center (HEASARC), provided by NASA's Goddard Space Flight Center.
Support for this work was provided by NASA through \Chandra\ Award Number
GO5-6020X issued by the {\it Chandra X-ray Observatory\/} Center, which is
operated by the Smithsonian Astrophysical Observatory for and on behalf of
NASA under contract NAS8-03060. This work was performed under the auspices of
the U.S.\ Department of Energy by University of California, Lawrence Livermore
National Laboratory under Contract W-7405-Eng-48.

\bsp

\label{lastpage}

\end{document}